\documentclass[journal]{IEEEtran}

\usepackage{xcolor,soul,framed} 

\colorlet{shadecolor}{yellow}
\usepackage[pdftex]{graphicx}
\graphicspath{{../pdf/}{../jpeg/}}
\DeclareGraphicsExtensions{.pdf,.jpeg,.png}

\usepackage[cmex10]{amsmath}
\usepackage{array}
\usepackage{mdwmath}
\usepackage{mdwtab}
\usepackage{eqparbox}
\usepackage{url}

\usepackage{pgf,interval}
\usepackage{graphicx}
\usepackage{xcolor}
\usepackage{interval}
\usepackage{amsmath, amssymb}
\usepackage{cuted, tcolorbox}

\usepackage{tikz}
\usepackage[ruled,vlined]{algorithm2e}
\usepackage{algpseudocode}

\newtheorem{definition}{Definition}

\newcommand{\VDFSETUP}{\mathsf{VDFSetup}}
\newcommand{\VDFEVAL}{\mathsf{VDFEval}}
\newcommand{\VDFVERIFY}{\mathsf{VDFVerify}}

\tikzset{
    state/.style={
           rectangle,
           rounded corners,
           draw=black, very thick,
           minimum height=2em,
           inner sep=2pt,
           text centered,
           },
}
\tikzstyle{pinstyle} = [pin edge={to-,thin,black}]

\begin{document}
\bstctlcite{IEEEexample:BSTcontrol}
    \title{Fair Proof-of-Stake using VDF+VRF Consensus }
  \author{Jos\'e~I.~Orlicki~\IEEEmembership{}
  }

\markboth{DRAFT DRAFT DRAFT DRAFT DRAFT DRAFT DRAFT }{}

\maketitle

\begin{abstract}
We propose a new Proof-of-Stake consensus protocol constructed with a verifiable random function (VRF) and a verifiable delay function (VDF) that has the following properties: a) all addresses with positive stake can participate; b) is fair because the coin stake is proportional to the distribution of rewards; c) is resistant to several classic blockchain attacks such as Sybil attacks, "Nothing-at-stake" attacks and "Winner-takes-all" attacks. We call it Vixify Consensus.
\end{abstract}


%
\IEEEpeerreviewmaketitle


\section{Introduction}

\label{sec:intro}

One of the most interesting abstract properties of Nakamoto distributed consensus is that this algorithm is very similar to implementing random clocks (with time inversely proportional to computing power) for the miners with the first clock to stop determining the block proposer. If we can implement this property in a non-parallelizable version we will save a lot of energy wasted on traditional Proof-of-Work and we might open the door also to Fair Proof-of-Stake with an unlimited number of block proposers.

A robust blockchain avoids centralization of stake and mining power, then we want to design a distributed consensus that discourages centralized stake or mining pools,  hardware parallelization, energy waste and Sybil-attacks \cite{sybil:02, bitcoin:09}. Also, classic Proof of Stake attack such as \emph{nothing-at-stake} attacks \cite{pos-attacks:19}.

The design objectives of this blockchain consensus for Proof of Stake \cite{protocols:19} are:

\begin{enumerate}
    \item\label{obj0} \emph{Mining-is-validating}: similarly to Proof-of-Work regarding transaction validation. While mining new blocks nodes are also validating transactions as the same time. Nodes work as both block proposers and validators.
    \item\label{obj1} \emph{Stake-aligned}: Complete alignment of stake distribution with rewards distribution during consensus ($Rewards(Stake) = 0$ for $Stake=0$).
    \item\label{obj2}\label{indep-aggr} \emph{Independent Aggregation}: Aggregating or separating stake into one or multiple accounts does not change the reward size ( $Rewards(S_1 + S_1) \approx Rewards(S_1) + Rewards(S_2) $ ). This property can be separated into the two better known properties following.
        \begin{itemize}
			\item\label{obj3} \emph{Sybil-tolerant}: Not susceptible to Sybil attacks, miners spawning multiple parallel block proposers ( $Rewards(S_1 + S_1) >= Rewards(S_1) + Rewards(S_2) $ ).
			\item\label{obj4} \emph{Pool-neutral}: Aggregating stake into pools does not provide any advantage ( $Rewards(S_1 + S_1) <= Rewards(S_1) + Rewards(S_2) $).
        \end{itemize}
    \item\label{obj5a} \emph{Consensus-scalability}: The consensus remains secure with arbitrary number of nodes.
    \item\label{obj5b} \emph{Permissionless}: Any node can join or leave the consensus at any time.
    \item\label{obj6} \emph{Fair Mining}: For any mining hardware, its mining speed eventually converges to a single predefined value.
     \item\label{obj7} \emph{Unbiased}: No adversary can manipulate who generates the next block, even equipped with powerful hardware \cite{leader:20}.
    \item\label{obj8} \emph{Unpredictable}: The probability that the adversary makes an accurate guess on the next block proposer is in proportion to the guessed node’s voting power. The more economic way to predict the next block winner given some stake, is to mine it \cite{leader:20}.
    \item\label{obj9} \emph{Fair Rewards}: Once a miner mines a block, its mining reward is in proportion to its stake.
\end{enumerate}

One the most important properties is number \ref{indep-aggr} because this property implies that the mining computation cannot be parallelized, i.e. is \emph{non-parallelizable}.

\section{Basic definitions}
\label{sec:defs}

\subsection{Verifiable random functions}
Verifiable Random Functions (VRFs) are now common, such as the one for Elliptic Curve $secp256k1$, a new standard for VRFs \cite{goldberg:19}, and are defined using a public-key pair $sk,pk$ having the property that using a private key $sk$ allows to hash a plain-text $s$ into a an hash $h$ that can be verified using a public key $pk$. VRFs are being popularized these days by the Algorand Blockchain project, although their consensus use a voting committee selected by VRF pseudo-randomness, instead of VDF mining.

VRF syntax and properties follows \cite{goldberg:19}. 

A VRF is a triple of algorithms $\mathsf{VRFkeygen}$, $\mathsf{VRFeval}$, and $\mathsf{VRFverify}$:

\begin{itemize}
    \item $\mathsf{VRFKeyGen}(r) \to (pk, sk)$. On a random input, the key generation algorithm produces a verification key $pk$ and a secret key $sk$ pair.
    \item $\mathsf{VRFEval}(sk, x) \to (h, \pi)$. The evaluation algorithm takes as input the secret key $sk$, a message $x$ and produces a pseudorandom output string $h$ and a proof $\pi$.
    \item $\mathsf{VRFVerify}(pk, x, h, \pi) \to \{0,1\}$. The verification algorithm takes as input the verification key $pk$, the message $x$, the output $h$ and the proof $\pi$. It outputs 1 if and only if it verifies that pseudo-random output $h$ is the output produced by the evaluation algorithm on inputs $sk$ and $x$.
\end{itemize}

VRF functions should satisfy also the properties \emph{VRF-Uniqueness}, \emph{VRF-Collision-Resistance} and \emph{VRF-Pseudorandomness}.
In a few words, VRF functions are public-key signing schemes where the signature is unique and pseudo-random.

Because pseudo-random $h$ outputs can be interpreted as fixed-size integer we can also generate more narrow range integer with modulo. Including integer $i$ as an upper bound:

We define $\mathsf{VRFEvalInt}(sk, x, n) \to y_{Int}$ as 
\begin{align}
    h, \pi \gets \mathsf{VRFEval}(sk, x)\\
    y_{Int} \gets h \mod n
\end{align}

\subsection{Verifiable Delay Functions}

Verifiable delays functions (VDFs) such as $\mathsf{VDFStep}(x,t)$ are essentials cryptographic hash functions computing $t$ steps of computation that cannot be parallelized but the computation can be verified much faster, or very fast \cite{vdfs:18}.
They have been proposed as solution to energy inefficient parallelizable Proof-of-Work consensus because of their non-parallelizable properties but they raised some concerns regarding "winner-takes-all" scenarios for nodes with very fast specialized hardware, such as ASIC hardware.

\begin{definition}[Verifiable Delay Function\cite{vdfs:18}]
    A Verifiable Delay Function is a tuple of three algorithms $(\VDFSETUP, \VDFEVAL, \VDFVERIFY)$ that satisfied the following properties\cite{vdfs:18}:
    \begin{itemize}
        \item \emph{Sequential}: honest parties can compute $\VDFEVAL(pp, x) = (y, \pi)$ in $t$ sequential steps, while no parallel-machine adversary with a polynomial number of processors can distinguish the output $y$ from random in significantly fewer steps.
        \item \emph{Efficiently verifiable}: We prefer $\VDFVERIFY$ to be as fast as possible for honest parties to compute; we require it to take total time $\mathcal{O}(polylog(t))$.
        \item \emph{Unique}: for all inputs x, it is difficulty to find a y for which $\VDFVERIFY(pp, x, y, \pi) = Yes$, but $\VDFEVAL(pp, x) \neq y$.
    \end{itemize}
\end{definition}

Practical VDFs with current working implementations were described by Pietrzak and Wesolowski \cite{vdf1:18,vdf2:19,vdf-impl:18}. Also a pseudo-VDF that does not scales shows and interesting asymmetric between proving and validating the proof is Sloth \cite{vdfs:18}.

\section{Consensus}
\label{sec:cons}

Although we discuss a non-prefixed number of steps VDF puzzle in this section (Algorithm \ref{puzzle-cont}), the final consensus proposed is not search for an output bigger than a difficult but just generated a VDF proof using a pre-determined number of steps cased on a VRF proof, i.e. a personalized random seed. Then we are most interested in the case of VDF mining based on VRF-randomized number of steps.

\subsection{Block structure}
\label{sec:block}

A special block structure with subblock independant of VDF outputs and inputs. We seggregate the fields that are dependent on the Merkle tree root chosen by the miner so that these fields are not inputs of the linear puzzles that decide who is the block proposer. This way any miner cannot control Merkle tree root to generate many parallel copies of linear VDF mining and increase its chances of being selected as block proposer.

\scalebox{0.80}{
\begin{tikzpicture}[->]

 \node[state,
    text width=4cm] (PREVBLOCK) 
 {\begin{tabular}{l}
 
  \textbf{Block(N-1)}\\ \\
  \emph{Txs. Indep. Sect.} (A):\\
  \parbox{4cm}{\begin{itemize}
   \item BlockHashA(N-2)
   \item MinerAddress(N-1)
   \item VRFProof(N-1)
  \end{itemize}
  }\\
    \emph{Txs. Dep. Sect.} (B):\\
  \parbox{4cm}{\begin{itemize}
   \item BlockHashA+B(N-2)
   \item VDFProof(N-1)
   \item MerkleRoot(N-1) 
   \item VRFProofMR(N-1)
  \end{itemize}
  }\\[4em]
 \end{tabular}};
  
 \node[state,    	
  text width=4cm, 	
  yshift=0cm, 		
  right of=PREVBLOCK, 	
  node distance=6.5cm, 	
  anchor=center] (BLOCK) 	
 {%
 \begin{tabular}{l} 	
 
 \textbf{Block(N)}\\ \\
  \emph{Payload Indep. Sect.} (A):\\
  \parbox{4cm}{\begin{itemize}
   \item BlockHashA(N-1)
   \item MinerAddress(N)
   \item VRFProof(N)
  \end{itemize}
  }\\
    \emph{Payload Dep. Sect.} (B):\\
  \parbox{4cm}{\begin{itemize}
   \item BlockHashA+B(N-1)
   \item VDFProof(N)
   \item MerkleRoot(N) 
    \item VRFProofMR(N)

  \end{itemize}
  }\\[4em]

 \end{tabular}
 };

 \path (PREVBLOCK) 	edge[bend left=10]  node[anchor=south,above]{Mine Next}
                                    node[anchor=north,below]{Block} (BLOCK);

\end{tikzpicture}
}

\subsection{System Setting}
\label{sec:setting}

Let $sk_k, pk_k$ be the pair of secret key and public key of the $k$-th miner.
Let $S_k$ be the stake fraction of miner $k$.
Let $N$ be the height of the latest block, and $T_N$ be the difficulty when the latest block is $N$.
Let $H'_{N-1}$ be the hash of $(N-1)$-th block's metadata, i.e., everything except for Merkle root.
Let $H(\cdot)$ be a hash function.

\subsection{Linear Puzzles}
\label{sec:puzzles}

We use linear puzzles to simulate Proof-of-Work \cite{bitcoin:09} but with a VDF.

\begin{algorithm}[]\label{puzzle0}
    \caption{Proof-of-Work puzzle from miner $k$'s perspective \cite{bitcoin:09}.} 
    \SetAlgoLined\DontPrintSemicolon
    \KwIn{}
    \KwOut{}
    $t \gets 0$\;
    \While{True}{
        $hash_{N,k}(t) \gets H(merkle_{N, k}, H_{N-1}, t)$\;
        \If{$hash_{N,k}(t) > T_N$}{
            $\mathsf{Nonce}_{N,k} \gets t$\;
            \Return{$\mathsf{Nonce}_{N,k}$}\;
        }
        $t += 1$\;
    }
\end{algorithm}

\begin{algorithm}[]\label{puzzle-vixify}
    \caption{Vixify puzzle with pre-calculated number of VDF steps (no need for continuous VDF in this case)}
    \SetAlgoLined\DontPrintSemicolon
    \KwIn{}
    \KwOut{}
    $in^{VRF}_{N} \gets HashAB({N-1})\oplus MerkleRoot(N)$\;
    $in^{VDF}_{N,k} \gets \mathsf{VRFEval}(sk_k, in^{VRF}_{N})$\;
    $\mathsf{Hash}_{N,k}(t) \gets \VDFEVAL(in^{VDF}_{N,k}, t)$ for $t\geq 0$\;
    $\mathsf{Range}_k \gets \lfloor \frac{1}{S_k} \rfloor$\;
    $\mathsf{Slot}_{N,k} \gets \mathsf{VRFEval}(sk_k, H'_{N-1}) \mod \mathsf{Range}_k$\;
        $\mathsf{Nonce}_{N,k} \gets \mathsf{Hash}_{N,k}( Q_N R_N^{\mathsf{Slot}_{N,k}} )  
    $
\end{algorithm}

\begin{algorithm}[]\label{puzzle-cont}
    \caption{Vixify puzzle with continuous VDF.}
    \SetAlgoLined\DontPrintSemicolon
    \KwIn{}
    \KwOut{}
    $\mathsf{VDFInput}_{N,k} \gets \mathsf{VRFEval}(sk_k, HashA(N-1))$\;
    $\mathsf{Hash}_{N,k}(t) \gets \mathsf{VDFStep}(\mathsf{VDFInput}_{N,k},t)$\;
    $\mathsf{Range}_k \gets \lfloor 1/S_k \rfloor$\;
    $\mathsf{RandomSlot}_{N,k} \gets \mathsf{VRFEvalInt}(sk_k, H'_{N-1},\mathsf{Range}_k)$\;
    $\mathsf{Nonce}_{N,k} \gets min \{\ t : \mathsf{Hash}_{N,k}(t) > Q_N R_N^{\mathsf{RandomSlot}_{N,k}} \ \}$
\end{algorithm}

From the binary point of view, we are searching for a $\mathsf{VDFStep}()$ output with a number of leading $1$s plus other binary conditions on the rest of the bits.
Each block proposer is mining $\mathsf{VDFStep}()$ until they find the first $\mathsf{Nonce}$ big enough to satisfy the conditions including the $\mathsf{RandomSlot}$.

So, current block proposer will be the one with the small number of steps satisfying their specific nonce restriction (there is no global condition for nonces):

$$
\mathsf{Proposer}_N \gets \{ k \in \mathsf{Miners} : \forall r \in\mathsf{Miners} : \mathsf{Nonce}_{N,k} \leq \mathsf{Nonce}_{N,r} \}
$$

Given the rare case there is more than one proposer per block, we can choose the one with the smaller number for steps. If there is also a collision on the number of steps then we can choose randomly based on $\mathsf{VDFInput}_{N,k}$, that is pseudorandom.

Difficulty is personalized for the stake of the miner and is discrete. For example, if the current pseudorandom discrete slot of the exponent is 0, then we have the linear mining for the first slot as: 

$$
min_{t} \{\ \mathsf{VDFStep}(\mathsf{VDFInput}_{N,k},t) > Q_N \ \}
$$

were $Q_N$ is a discrete quantum, same for all miners, that is dynamically adjusted each block based on average block time considering a large number of previous blocks.

Also $B_N$ is also the same for all miners on the current block, and allows an exponential adjustment that can protect the consensus from persistent strong hardware or strong optimized software attacks, miners with a VDF speed substantially faster than the rest of the miners.

\subsection{Dynamic Difficulty}
\label{sec:difficulty}

There is no question that stake fragmentation increases average block time because having many miners with little stake allows only more difficult puzzles than a small number of miners with large stake portions. If average block time get bigger we assume this is because the stake fragmentation has increased. Then is enough to reduce the difficulty linearly. The base quantum of VDF difficulty $Q_N$ is reduced by a very fraction $\alpha_N$ dynamic as per every block. This tendency can be reverted if stakes are being consolidated at some point, then we need increase linearly $Q_N$ (See Algorithm~\ref{difficulty1}). With these changes we want an stable average block time around a fixed number of seconds. 

\begin{algorithm}[]\label{difficulty1}
    \caption{Vixify difficulty adjustment for average block time. Similar to traditional Proof-of-Work.}
    \SetAlgoLined\DontPrintSemicolon
    \KwIn{$N$ block number, $Q_N$ current block time difficulty, $\alpha$ fractional change per block, $A^0$ target block time, $A$ current moving windows average block time for fixed windows size $a$}
    \KwOut{$Q_{N+1}$}
    
    \eIf{ $A >= A^0$}{
        $Q_{N+1} \gets Q_N * (1 - \alpha)  $\;
   }{
        $Q_{N+1} \gets Q_N * (1 + \alpha)  $\;
  }
\end{algorithm}

Average block time is also influenced by VDF speed, average number of VDF steps per second. We need to account for that also using timestamps in blocks. If average VDF steps per second get smaller we assume that this is because hardware or software optimizations have made the VDF computation faster. This tendency can be economically reverted if big stakeholders sell their stake and stop mining, then is bi-directional. Then we can use the maximum speed to date as the reference.  Then the algorithmic adjustment for this will be exponential on the exponential base $R_N$ only in the fractional increment $\beta_N$ if the maximum speed has been surpassed (see Algorithm~\ref{difficulty2}). Because there is not target for VDF speed then we must choose a big moving average window $b$ for VDF speed and a small change fraction $\beta$ so this difficulty $R_N$ moves very slowly. This allows miners investing in faster VDF hardware or software profitable for small time span like minutes or hours, allowing a miner $k$ on block $N$ to jump from one random slot $\mathsf{VRFEvalInt}(sk_k, H'_{N-1},\lfloor 1/S_k \rfloor)$ to a smaller one only for such a small time.

\begin{algorithm}[]\label{difficulty2}
    \caption{Vixify difficulty adjustment for average VDF steps per second. Affects slots exponentially to quickly reduce optimization advantage of any miner.}
    \SetAlgoLined\DontPrintSemicolon
    \KwIn{$N$ block number, $R_N$ current block VDF speed difficulty, $\beta$ fractional change per block, $B_N$ current moving windows average VDF steps per second for fixed windows size $b$}
    \KwOut{$B_{N+1}$}
    
    \eIf{ $B_N >= B_{N-1}$}{
        $R_{N+1} \gets R_N * (1 - \beta)  $\;
   }{
        $R_{N+1} \gets R_N * (1 + \beta)  $\;
  }
\end{algorithm}

Block timestamps can be manipulated by miners but in a very limited way. If they lie and produce bigger timestamps, other peers will detect that and will not propagate those blocks, if the produce smaller timestamps it will increase the difficulty then making mining harder for everyone including themselves. Same with VDF speed, because the only slack for miners is distorting timestamps (they cannot distort the VDF number of steps) without risking loosing the opportunity to propose a winning block. 



\begin{algorithm}[]\label{puzzle-beacon}
    \caption{Vixify linear puzzle with pre-calculated number of VDF steps (no need for continuous VDF in this case)}
    \SetAlgoLined\DontPrintSemicolon
    \KwIn{}
    \KwOut{}
    $in^{VDF}_{N,k} \gets \mathsf{VRFEval}(sk_k, HashAB({N-1}))$\;
    $\mathsf{Hash}_{N,k}(t) \gets \VDFEVAL(in^{VDF}_{N,k}, t)$ for $t\geq 0$\;
    $\mathsf{Range}_k \gets \lfloor \frac{1}{S_k} \rfloor$\;
    $\mathsf{Slot}_{N,k} \gets \mathsf{VRFEval}(sk_k, H'_{N-1}) \mod \mathsf{Range}_k$\;
        $\mathsf{Nonce}_{N,k} \gets \mathsf{Hash}_{N,k}( Q_N R_N^{\mathsf{Slot}_{N,k}} )  
    $
\end{algorithm}

\section{Verification}
\label{sec:verification}

\subsection{Mining-is-validating}

\begin{proof}
The Merkle-tree root hash is a cryptography digest of the transactions payload. This hash is included as part of the input of the VDF function for each block and for each miner, see Algorithm \ref{puzzle-vixify}. This proves that the each valid block proposed to be accepted must be a block with a valid set of transactions.
\end{proof}

\subsection{Stake-aligned}

\begin{proof}
This is trivially true because it can be checked that if the stake of the miner is zero it cannot propose any valid block. Otherwise we will find a divided-by-zero error when calculating the miner slot $1/S_k$.
\end{proof}

\subsection{Independent Aggregation}

The most important property for this Proof-of-Stake consensus is Independent Aggregations (i.e. of stake). The proof of this Property \ref{indep-aggr} of Proof-of-Stake (Section \ref{sec:intro}) can be sketched in the following way:

\begin{proof}

\begin{enumerate}
    \item Prove Property \ref{indep-aggr} for stakes of the form $1/2^k$ that are split into two ($\geq$, Sybil-tolerant).
    \item Prove Property \ref{indep-aggr} for stakes of the form $1/2^k$ that are aggregated ($\leq$, Pool-neutral).
    \item Prove Property \ref{indep-aggr} for all stake because they have the form $\Sigma_{k\in } 1/2^k$
\end{enumerate}

\end{proof}

\subsection{Consensus-scalability and Permission-less}

\begin{proof}
Our distributed consensus is very similar to Nakamoto consensus so these properties are directly satisfied.

We are not using a Voting Committee as other protocol, then, there is not limit to the number of miners proposing blocks as long as they have a positive balance of coins, also know as stake.  

Any fraction of nodes with any given fraction of stake can leave the consensus at any time and the protocol is robust to continue operating with a block time that will be bigger for some time.
\end{proof}

\subsection{Fair Mining}

This property is proved based on the dynamic exponential difficulty adjustment based on the current mean VDF speed (see Algorithm \ref{difficulty2}).

\subsection{Unbiased and Unpredictable}

\begin{proof}
Sketch: is very similar to Nakamoto Consensus but in this case the probability of being the first to propose a block and win is proportional to the stake of the miner.
\end{proof}

\subsection{Fair Rewards}

Block proposer VDF difficulty (number of VDF steps) is based on slots $1/S_k$ determined by stake $S_k$. This is designed to be a very good approximation of stake itself.

\section{Security Analysis}
\label{sec:security}

\subsection{Nothing-at-Stake}
\label{sec:nothing}

Miners cannot generate new addresses to do mining on each block, because each new address requires a number of staked coins on the address balance, then due to Property \ref{indep-aggr} of Proof-of-Stake (Section \ref{sec:intro}) there is not rewards gain in splitting the stake into several addresses.

\subsection{Winner-takes-all}
\label{sec:winner}

The exponential difficulty parameter (Algorithm \ref{puzzle-vixify}) and its slow but steady adjustment (Section \ref{difficulty2}) makes software or hardware optimizations of VDF speed only profitable for a short time. After this short time of adjustment is very difficult to jump from the assigned slot to a smaller one.

\subsection{Implementation}
\label{sec:impl}

An proof-of-concept was implemented to test the feasibility of the design, excluding the block data segregation protections for Sybil-attacks. The implementation uses RSA for VRF generation and blockchain signatures, and on the VDF we are using a popular pseudo-VRF called Sloth\cite{impl}. Slot is pseudo-VDF because the the verification takes only a fixed fraction of the evaluation time, around 40 times.

\section{Conclusion}
\label{sec:conc}

In summary, this paper addresses using a Sequential Proof-of-Work Consensus, called Vixify. It is based on a verifiable delay function (VDF) and verifiable random function (VRF) to simulate a distributed consensus very similar to Nakamoto Consensus but energy-efficient, fair with stakes, and resistant to Sybil attacks and hardware optimizations.

This distributed consensus we proposed has satisfied the abstract property of Nakamoto Consensus of simulating random clocks running on each miner but without the possibility of \emph{parallelizing} the computation, then the timer for each miner is not inversely proportional to their computing power but directly proportional to their stake in coins. 

\section*{Acknowledgment}

The authors would like to thank Santiago Bazerque from HyperHyperSpace for interesting conversations on Blockchain Scaling and Sharding that motivated the formalization of this distributed consensus algorithm.


%





\ifCLASSOPTIONcaptionsoff
  \newpage
\fi





\bibliographystyle{IEEEtran}
\bibliography{IEEEabrv,Bibliography}
%

\begin{IEEEbiography}[{\includegraphics[width=1in,height=1.25in,clip,keepaspectratio]{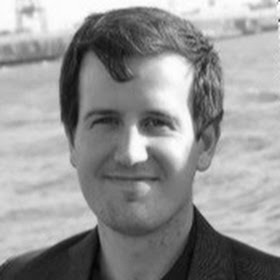}}]{Jos\'e I. Orlicki}
 received the 6-year B.S.C.S degree from Universidad de Buenos Aires, Buenos Aires, CABA, in 2006, and a MSc in Financial Engineering degree from the Stevens Institute of Technology, Hoboken, NJ, in 2017. From 2006 to 2009, he was a Computer Security Researcher with Core Security Technologies, Buenos Aires, CABA, where he was involved with network attack simulations and automated attack planning. Hi also worked as Machine Learning Specialist for 7Puentes for several years, involved in Supervised Learning, Recommender Systems and Web Scraping applications. His current research interests include algorithmic trading, cryptoeconomics, stablecoins and distributed consensus. He has collaborated in many active blockchain projects.
\end{IEEEbiography}





\vfill


\end{document}